\documentclass[acmtog, authorversion]{acmart}

\usepackage{amsmath}

\DeclareMathOperator*{\argmin}{argmin}
\usepackage{bm}
\usepackage{cleveref}
\usepackage{multirow}
\usepackage{float}

\copyrightyear{2026}
\acmYear{2026}
\setcopyright{cc}
\setcctype{by}
\acmConference[SA Conference Papers '26]{SIGGRAPH Asia 2026 Conference Papers}{December 01--04, 2026}{Kuala Lumpur, Malaysia}
\acmBooktitle{SIGGRAPH Asia 2026 Conference Papers (SA Conference Papers '26), December 01--04, 2026, Kuala Lumpur, Malaysia}
\acmDOI{10.1145/3829340.3842376}
\acmISBN{979-8-4007-2842-6/2026/12}

\acmSubmissionID{2752}

\begin{document}

\title{RealSimLoop: Online Real-to-Sim Adaptation via Differentiable Reduced-Order Simulation with Vision Feedback}

\author{Zhihao Cen}
\email{czh1224415633@gmail.com}
\orcid{0009-0009-5967-3895}
\affiliation{
  \institution{South China University of Technology}
  \city{Guangzhou}
  \country{China}
}

\author{Chuhua Xian}
\authornote{Corresponding authors.}
\email{chhxian@scut.edu.cn}
\orcid{0000-0001-7656-4652}
\affiliation{
  \institution{South China University of Technology}
  \city{Guangzhou}
  \country{China}
}

\author{Hailin Sun}
\email{hlsun@mae.cuhk.edu.hk}
\orcid{0000-0002-7251-2860}
\affiliation{
  \institution{The Chinese University of Hong Kong}
  \city{Hong Kong SAR}
  \country{China}
}

\author{Yuliang Liufu}
\email{yuliangliufu@cuhk.edu.hk}
\orcid{0009-0007-4463-0337}
\affiliation{
  \institution{The Chinese University of Hong Kong}
  \city{Hong Kong SAR}
  \country{China}
}

\author{Zhen Zhang}
\email{zhzhen@link.cuhk.edu.hk}
\orcid{0009-0003-9587-6864}
\affiliation{
  \institution{The Chinese University of Hong Kong}
  \city{Hong Kong SAR}
  \country{China}
}

\author{Xiangyu Chu}
\email{xiangyuchu@cuhk.edu.hk}
\orcid{0000-0002-7677-2600}
\affiliation{
  \institution{The Chinese University of Hong Kong / Multi-scale
Medical Robotics Centre, Hong Kong SAR}
  \city{Hong Kong SAR}
  \country{China}
}

\author{Hongmin Cai}
\email{hmcai@scut.edu.cn}
\orcid{0000-0002-2747-7234}
\affiliation{
  \institution{South China University of Technology}
  \city{Guangzhou}
  \country{China}
}

\author{Yunbo Zhang}
\email{yunbozhang@hkust-gz.edu.cn}
\orcid{0000-0003-0254-7168}
\affiliation{
  \institution{Hong Kong Institute of Science \& Innovation, CAS, Hong Kong}
  \city{Hong Kong}
  \country{China}
}

\author{Guoxin Fang}
\authornotemark[1]
\email{guoxinfang@cuhk.edu.hk}
\orcid{0000-0001-8741-3227}
\affiliation{
  \institution{The Chinese University of Hong Kong}
  \city{Hong Kong SAR}
  \country{China}
}

\newcommand{\guoxin}[1]{\textcolor{blue}{\textit{[Guoxin:~#1]}}}

\newcommand{\rev}[2]{\textcolor{black}{#2}}

\newcommand{\aoran}[1]{\textcolor{red}{\textit{[Aoran:~#1]}}}
\newcommand{\zhihao}[1]{\textcolor{green}{\textit{[Zhihao:~#1]}}}
\newcommand{\xian}[1]{\textcolor{pink}{\textit{[XIAN:~#1]}}}
\newcommand{\js}[1]{\textcolor{brown}{\textit{[Js:~#1]}}}

\begin{abstract}

Real-world observations of deformable objects are often sparse or surface-level, while downstream tasks require hidden physical quantities such as internal deformation, stress fields, and interaction forces. Physics-based simulation can recover these quantities, but online real-to-sim adaptation remains challenging due to costly full-space optimization, limited feedback, and time-varying material properties. To address these challenges, we propose RealSimLoop, a differentiable framework for online real-to-sim adaptation using vision data as physical feedback. Our approach achieves quasi-real-time performance by executing differentiable simulation within a reduced-order neural subspace, drastically accelerating the optimization loop. We couple this efficient dynamics model with differentiable rendering, enabling direct gradient backpropagation that leverages high-fidelity pixel data to refine physical parameters such as material stiffness. Furthermore, by employing a sliding-window objective function, RealSimLoop enables robust online adaptation, allowing the system to track time-varying material properties and effectively bridge the real-to-sim gap arising from model reduction or unmodeled dynamics. Extensive experiments demonstrate that our method outperforms conventional offline methods, and we validate the framework's versatility in downstream applications, including external force prediction and 3D stress field reconstruction with novel view synthesis.

\end{abstract}


\begin{teaserfigure}
  \includegraphics[width=\textwidth]{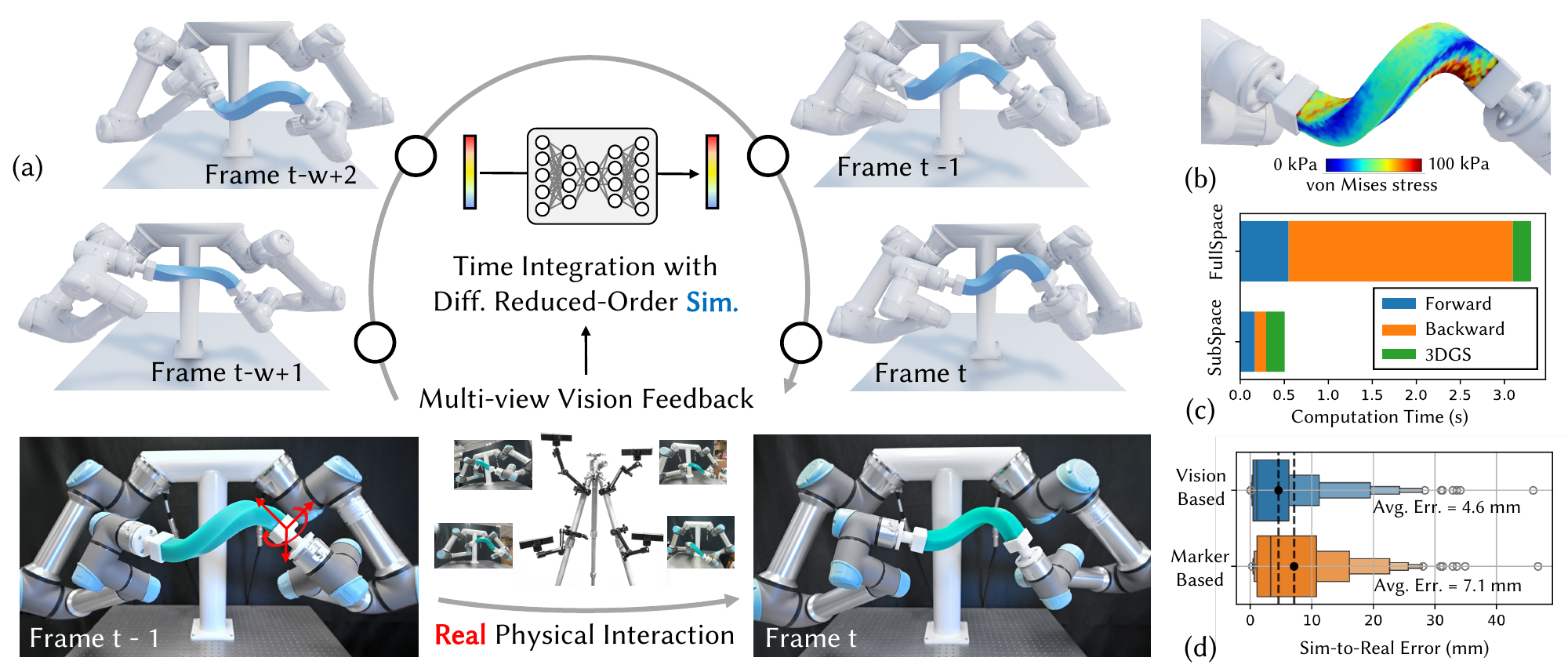}
  \caption{
  We propose an \textit{online} Real-to-Sim method to reduce the reality gap for deformable object manipulation. (a) In our pipeline, real-to-sim gap is eliminated by minimizing the difference between physical vision feedback and simulation-based rendered images within a sliding time \textit{window} $\{t - w + 1,\dots,t\}$. In this way, we close the sim-real loop by integrating scene representation with 3D-GS, (b) and realize downstream applications like physical stress construction in arbitrary new synthesized views. (c) Neural reduced-order differentiable simulation is adopted, which significantly accelerates computation and achieves quasi-real-time performance, delivering 6.23x speedup over full-space methods. (d) Our method generalizes to physical inputs from conventional marker-based motion capture systems, whereas the vision-based solution performs better. 
}
  \label{fig:teaser}
\end{teaserfigure}

\maketitle






\section{Introduction}
\label{sec:intro}

Bridging the reality gap between the physical world and digital simulation is a cornerstone of digital twin technologies, enabling a wide range of applications, including robotic manipulation of deformable objects (Fig.~\ref{fig:teaser}), multi-material structural analysis (Fig.~\ref{fig:tower}), \rev{}{ heat-sensitive stiffness tracking (Fig.~\ref{fig:gingerbread-real}), and structural health monitoring (Fig.~\ref{fig-bar-appendix}). } In these applications, real-world observations are often limited to sparse or surface-level geometric measurements captured by motion capture or vision systems. Physics-based simulation can complement such observations by inferring hidden physical quantities, such as internal deformation, stress distribution, and interaction forces~\cite{gjoka2024soft, kim2017data}. This makes real-to-sim adaptation essential for keeping simulations aligned with real-world physical behavior~\cite{nealen2006physically}.

However, achieving this level of physical understanding remains challenging. Conventional Real-to-Sim adaptation methods typically conduct an \textit{offline} solution~\cite{cai2024gic}: collecting a set of pre-recorded datasets, then fitting simulation parameters using parametric material models (e.g., linear elasticity, Saint Venant-Kirchhoff (StVK) model~\cite{barbivc2005real}, Neo-Hookean model~\cite{smith2018stable}, etc.). However, this may introduce significant modeling errors - particularly for hyperelastic materials, as idealized constitutive laws rarely match real-world complex laws perfectly. On the other hand, these offline approaches can suffer from open-loop drift, resulting in a lack of adaptability to time-varying properties or sudden environmental changes~\cite{hahn2019real2sim} (example also shown in Fig.~\ref{fig-gingerbread-show} and discussed in Sec.~\ref{subsec:result}).

Recently, data-driven approaches based on vision systems have enabled online scene reconstruction~\cite{wu20244d}. Although these advances enable the effective recovery of time-varying surface geometry from video inputs, they suffer from a fundamental disconnect with physical reality~\cite{wang2025vggt}. Critical mechanical information for downstream applications, such as the prediction of internal stress fields, force boundary conditions, and material properties (see the application demonstrated in Sec.~\ref{subsec:result}), cannot be inferred from these approaches.

\subsection{Our method and contribution}

In this work, we introduce \textit{RealSimLoop}, a novel quasi-real-time \textit{online} real-to-sim adaptation framework that continuously synchronizes a differentiable physical simulator with vision feedback. 
Our pipeline is driven by efficient scene representation (e.g., 3D Gaussian Splatting~\cite{kerbl20233d}), and coupled with a time \textit{window}-based objective function that allows us to capture dynamic variations in material properties over time \rev{}{to reduce the accumulated sim-to-real gap~\cite{liu2021real}. }
To bridge the gap between observed geometry and underlying physical parameters, we employ a differentiable simulation engine capable of highly efficient forward and backward steps. 
Our system also enables seamless online operation by integrating a neural-subspace reduced-order model~\cite{fulton2019latent}, which accelerates high-fidelity soft-body simulation. While such models are prone to approximation errors~\cite{shen2021high}, our framework leverages continuous visual feedback to dynamically compensate for these inaccuracies, therefore maintaining the efficiency to eliminate the real-to-sim gap.



\rev{}{\textit{RealSimLoop} addresses the system-level challenge of continuously maintaining physically meaningful deformable-object simulations by integrating visual feedback, efficient differentiable simulation, and online material adaptation under practical computational constraints.} Our technical contributions are summarized as follows:
\begin{itemize}

    \item We propose a sliding-window optimization approach for online real-to-sim adaptation, reducing open-loop drift and aligning simulation with evolving real-world behavior.
    \item We employ an efficient neural-based differentiable simulator that accelerates forward simulation and gradient backpropagation, making online gradient-based real-to-sim adaptation computationally feasible.

    \item We integrate differentiable simulation with differentiable rendering to incorporate vision-based physical feedback, enabling closed-loop adaptation from image observations.
\end{itemize}

We validate the proposed method in different virtual and physical settings. Compared with offline methods~\cite{cai2024gic, hahn2019real2sim} and purely vision-based pipelines~\cite{wang2025vggt, wu20244d}, our approach continuously closes the real-to-sim gap while preserving physical consistency. We showcase practical applications in robotic manipulation, force estimation, and structural analysis. \rev{}{The source code of this work will be released upon acceptance.}

\section{Related Work}
We review literature on differentiable simulation and Real-to-Sim adaptation for both graphics and robotics applications, and discuss recent advances in neural reduced-order simulation methods.

\begin{figure*}[!t]
  \centering
  \includegraphics[width=0.8\linewidth]{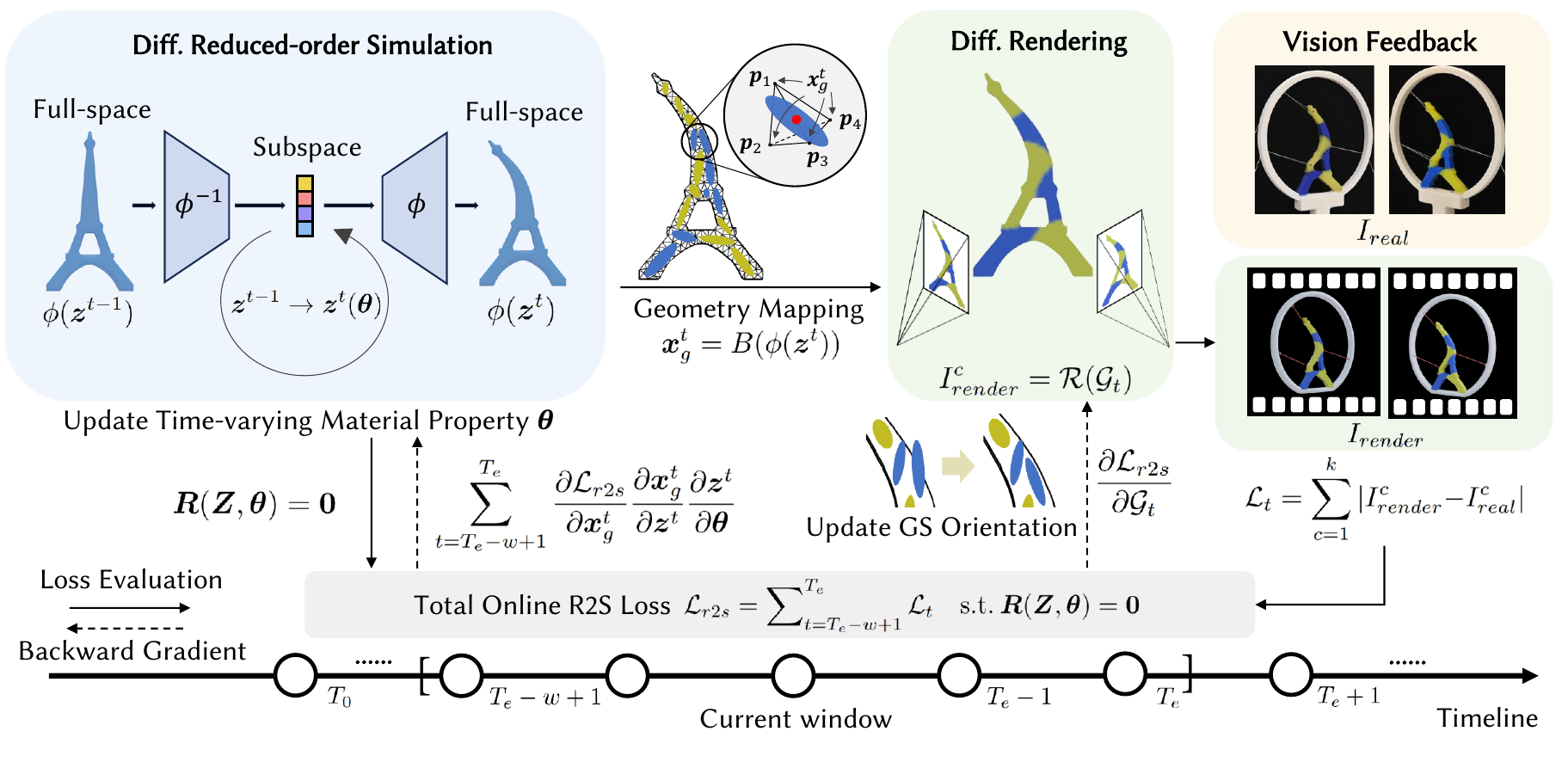}
  \vspace{-15px}
  \caption{Pipeline for \textit{online} real-to-sim adaptation with vision-based feedback (i.e., sparse-view images $I_{\text{real}}$ obtained from cameras). The forward process simulates deformation using a reduced-order model, maps the geometry to 3D Gaussians, and renders the image $I_{\text{render}}$ through differentiable rendering to evaluate $\mathcal{L}_{\text{s2r}}$ by updating time-varying material properties. The backward process 
  back-propagates loss gradients to update material properties $\bm{\theta}$ and Gaussian orientations within a sliding window (i.e., from $T_e$-w+1 to $T_e$), supporting gradient-based optimization to efficiently eliminate the real-sim difference.}
  \label{fig:pipeline}
\end{figure*}

\subsection{Real-to-sim adaptation from physical observations}

Aligning physics-based simulation with real-world observations is essential for digital twins, robotic manipulation, and physical state reconstruction. Existing solutions often rely on expensive data acquisition devices, such as 
motion capture systems~\cite{hahn2019real2sim, chen2017dynamics} and specialized probing or force-sensing setups~\cite{pai2001scanning, bickel2009capture, pai2018human}, to capture the physical world and calibrate simulations. Image-based observations offer a scalable alternative: prior work estimates cloth material properties from images~\cite{bouman2013estimating, davis2015visual, yang2017learning}, while recent differentiable rendering methods enable image-level losses to optimize simulation states and physical parameters~\cite{jatavallabhula2021gradsim, cai2024gic, rho2025projo4d}. \rev{}{Recent works~\cite{modi2024simplicits, zhong2024reconstruction, jiang2025phystwin, li2023pac} have also invited vision data to guide the real-to-sim adaptation with deformable objects.} Nevertheless, most methods remain \textit{offline}, i.e., fitting a single set of parameters to a pre-collected sequence, and therefore struggle to track evolving physical properties during dynamic interactions.

A major barrier to online real-to-sim adaptation is the high computational cost of repeatedly simulating and differentiating physical dynamics. Although recent methods incorporate the Material Point Method (MPM)~\cite{hu2018moving, xie2024physgaussian}, they often remain too expensive for real-time use. \rev{}{Similarly, Vid2Sim~\cite{chen2025vid2sim} employs a mesh-free simulation model to reduce complexity but remains unsuitable for online adaptation over long sequences.} To address this, we introduce a sliding-window adaptation objective integrated with reduced-order differentiable simulation, improving computational efficiency while enabling robust online tracking of evolving physical properties.

\subsection{Differentiable and Reduced-Order Simulation}

Differentiable simulation has become a vital research direction in the computer graphics community, enabling the computation of gradients for simulation outputs w.r.t. physical parameters and initial states~\cite{degrave2019differentiable, huang2024differentiable}. This capability facilitates gradient-based optimization for inverse problems, driving applications in rigid body control~\cite{xu2021end} and cloth manipulation~\cite{li2022diffcloth}. In particular, it has been widely introduced for soft body simulation~\cite{hu2019chainqueen, du2021diffpd}, where~\citeN{hahn2019real2sim} have demonstrated its ability for \textit{offline} real-to-sim transfer.

On the other hand, Reduced-order simulation has emerged as a key direction for accelerating physics by mapping high-dimensional systems to low-dimensional subspaces~\cite{chang2023licrom, sharp2023data}. While early linear methods like modal analysis~\cite{pentland1989good, james2002dyrt} and PCA~\cite{krysl2001dimensional} require large subspaces for nonlinear deformations, recent neural networks enable nonlinear mappings with superior representation capabilities~\cite{fulton2019latent}. Although techniques like Lipschitz loss~\cite{lyu2024accelerate} or linear corrections~\cite{shen2021high} optimize speed, they rely on supervised data; conversely, unsupervised alternatives~\cite{sharp2023data, wang2024neural} can suffer from physical drift. In this work, we leverage the differentiable reduced-order simulator with vision feedback to eliminate the real-to-sim gap, utilizing the neural-based nonlinear mapping from~\cite{fulton2019latent} to capture complex deformations.

\section{Preliminary and Overview}

In this section, we first provide preliminary knowledge on simulation and the formulation of the online real-to-sim adaptation problem. Then, we give a short discussion followed by an overview of our pipeline (illustrated in Fig.~\ref{fig:pipeline}).

\subsection{Preliminary of Simulation and Problem of Real-to-Sim}

Consider a deformable object discretized by a tetrahedral mesh $\mathcal{M}$ consisting of $n$ vertices and $m$ elements. The system configuration of the object at time $t$ is represented by a flattened vector $\bm{x}^t \in \mathbb{R}^{3n}$, obtained by stacking the positions of all vertices. For a dynamic system using implicit time integration~\cite{baraff2023large}, the forward deformable simulation can be formulated as:
\begin{equation}
\label{eq:forward_full}
\bm{r}(\bm{x}^{t},\bm{y}^{t}, \bm{\theta}):=\frac{1}{\Delta t^2} \bm{M} (\bm{x}^{t}-\bm{y}^{t}) - \bm{f}(\bm{x}^{t}, \bm{\theta}) = \bm{0},
\end{equation}
where $\Delta t$ is the timestep, $\boldsymbol{M}$ is the lumped mass matrix, $\bm{\theta} = [E, v]$ is the material parameters representing Young's modulus and Poisson's ratio respectively, $\bm{y}^{t}:=\bm{x}^{t-1}+\Delta t \bm{v}^{t-1}$, $\bm{v}^{t-1}=(\bm{x}^{t-1} - \bm{x}^{t-2})/\Delta t$, $\bm{f}(\bm{x}^{t}, \bm{\theta})$ denotes the total forces including elastic internal force, gravity, and interaction force. Consider a sequence of timesteps $t=0,1,\cdots,T$, forward simulation solves the nonlinear system:
\begin{equation}
\bm{r}(\bm{x}^{t},\bm{y}^{t}, \bm{\theta}) = \bm{0}, \forall t \in \{0,\dots, T\},
\end{equation}
As $\bm{y}^t$ is a function of $\bm{x}^{t-1}$ and $\bm{v}^{t-1}$, and $\bm{v}^{t-1}$ is either a boundary condition or a function of $\bm{x}^{t-1}$ and $\bm{x}^{t-2}$, we can stack the DoF vectors of all timesteps as a vector $\bm{X} \in \mathbb{R}^{T\times 3\times n}$. And the residual functions of the nonlinear system can be also stacked as a single function $\bm{R}(\bm{X}, \bm{\theta})$. Forward simulation then solves $\bm{X}$ such that the state equation $\bm{R}(\bm{X}, \bm{\theta})=0$ holds.

To eliminate the real-to-sim gap, an objective function $\mathcal{L}_{\text{r2s}}$ that describes the difference between the simulation result $\bm{X}$ and the observation (e.g., motion capture systems, multi-view RGB feedings) of the real-world experiment is introduced. In this work, we try to correct the reality gap by finding empirically the best material parameters $\bm{\theta}$ to minimize the discrepancy between simulation and real-world observation, which can be written as a constrained optimization problem:
\begin{equation}
\label{eq:SI_general}
\argmin_{\bm{\theta}}~\mathcal{L}_{\text{r2s}}(\bm{X}) \ \ \text{s.t.}~\bm{R}(\bm{X}, \bm{\theta})=\bm{0}.
\end{equation}
which is often solved by gradient-based optimization methods - details to be discussed in Sec.~\ref{subsec:optimization}. 

\subsection{Short Discussion and Overview}
\label{subsec:discuss}

In conventional \textit{offline} methods, the objective $\mathcal{L}_{\text{r2s}}$ is typically defined as the discrepancy between simulated trajectories and ground-truth geometry data collected via a motion capture system ~\cite{hahn2019real2sim} as:
\begin{equation}
\mathcal{L}_{\text{r2s}-mark}^{\text{off}}(\bm{\theta}) = \int_{t=0}^{T} \sum_{i=1}^k \|\mathbf{p}_{sim}^i(\bm{\theta}) - \mathbf{p}_{real}^i\|^2 dt,
\end{equation}
which aggregates sim-real error over time $T$ at marker points $\mathbf{p} \in \mathcal{M}$. While effective for small strains, the accuracy of offline methods will collapse at large elongations (>100\%), \rev{}{even with complex material models such as the Yeoh model~\cite{yeoh1993some} (see Fig.\ref{fig:dinosaur-multi-material} and Sec.\ref{sec:Dis} for a detailed discussion).} 
Meanwhile, constant material parameters (i.e., $\bm{\theta}$ kept unchanged across the entire time domain) cannot compensate for the error accumulation inherent in long-horizon forward simulations.

To handle this issue and greatly improve the accuracy of simulation, we propose the $\textit{online}$ pipeline (as illustrated in Fig.~\ref{fig:pipeline}), where the reality gap accumulation issue is handled through a \textit{window}-based real-to-sim objective with the update of material properties through time steps, which minimizes the vision-based loss accumulated over the current window $\{T_e - w + 1,\dots,T_e\}$ to update the material properties $\bm{\theta}$. Here $T_e$ is the current time step, $w$ is the window size.

To ensure the speed of online update, neural reduced-order solution is invited to greatly improve the simulation speed, where we introduce a subspace mapping $\phi$ obtained by data-driven method into the time-integration of forward simulation (we refer~\cite{fulton2019latent} for a comprehensive description of the neural reduced order simulations). Together with time-varying material properties $\bm{\theta}$, the time integration of forward simulation Eq.~\ref{eq:forward_full} is reformulated as
\begin{equation}
\label{eq:forward_sub}
\bm{J}(\bm{z}^{t})^\top\bm{r}(\bm{z}^{t}, \bm{y}^{t}, \bm{\theta}) := \bm{J}(\bm{z}^{t})^\top\big[\frac{1}{\Delta t^2}\bm{M}\big(\phi(\bm{z}^{t})-\bm{y}^{t}\big) - \bm{f}\big(\phi(\bm{z}^{t}), \bm{\theta} \big)\big]=\bm{0}
\end{equation}
where the subspace mapping function $\phi:\mathbb{R}^r \rightarrow \mathbb{R}^{3n}$ maps the low-dimensional subspace state $z \in \mathbb{R}^r$ to the high-dimensional full space (i.e., making $\bm{x}=\phi(\bm{z})$), $\bm{J}(\bm{z}^t) = \frac{\partial \bm{x}^t}{\partial \bm{z}^t} = \frac{\partial \phi}{\partial \bm{z}} \in \mathbb{R}^{3n \times r}$ is the Jacobian of the neural subspace mapping,
$\bm{y}^{t}:=\phi(\bm{z}^{t-1})+\Delta t \bm{v}^{t-1}$ and $\bm{v}^{t-1}=(\phi(\bm{z}^{t-1}) - \phi(\bm{z}^{t-2}))/\Delta t$. Similarly, we give the definition of the nonlinear system in the reduced subspace for the window time $w$:
\begin{equation}
\bm{J}(\bm{z}^{t})^\top\bm{r}(\bm{z}^{t},\bm{y}^{t}, \bm{\theta}) = \bm{0}, \forall t \in \{T_e - w + 1,\dots,T_e\},
\end{equation}
Further, the stacked DoF vectors across all timesteps can be rewritten as $\bm{Z} \in \mathbb{R}^{w\times r}$ and the residual functions of the nonlinear system as the constraint for online real-to-sim adaptation problem can also be formulated as $\bm{R}(\bm{Z}, \bm{\theta}) = \bm{0}$ (similar to the constraint in Eq.~\ref{eq:SI_general}).

On the other hand, unlike conventional approaches that rely solely on geometric information (i.e., marker positions captured by motion capture system), our framework leverages high-density visual data to capture complex deformation patterns and highly dynamic systems (see comparisons in Sec.~\ref{subsec:result}). To achieve high-efficiency geometry-to-image mapping, we integrate 3D Gaussian Splatting (3DGS) with differentiable rendering, thereby defining a differentiable \textit{online} objective with physical vision feedback to eliminate the reality gap. As illustrated by the dashed line in Fig.~\ref{fig:pipeline}, the differentiability of both simulation and rendering provides efficient gradient evaluation, allowing for robust optimization of the online real-to-sim adaptation and supporting downstream practical applications. 

\section{Method and Details}

We now detail the method applied to realize online sim-to-real adaptation with vision-feedback, starting from 3DGS-based mapping for an online window-based objective, then give the differentiable form of the reduced-order simulation to support gradient-based optimization.

\subsection{GS-Based Vision-Geometry Mapping}

In our pipeline, Gaussian primitives at timestep $t$ is defined as $\mathcal{G}_t=\{\bm{x}_g^t, \Sigma_g^t,\bm{c}_g^t,\sigma_g^t\}$. For these Gaussian attributes, $\bm{x}_g$, $\Sigma_g$, $\bm{c}_g$, and $\sigma_g$ denote the center position, covariance matrix, spherical harmonic coefficients, and opacity, respectively.
To establish a geometric mapping between the Gaussian points and the mesh, we express each Gaussian's position as a barycentric interpolation of its enclosing tetrahedron's vertices (i.e., $\bm{x}_g = \sum_{i=1}^4 w_i\bm{p}_i, s.t., \sum_{i=1}^4w_i = 1$, \rev{}{where $\bm{p}_i$ denotes the coordinate of the i-th vertex.}). Here we denote this geometry mapping function $B(\cdot)$ w.r.t. reduced-order simulation's latent space coordinates $\bm{z}^t$ (defined in Eq.~\ref{eq:forward_sub}) as $\bm{x}_g^t = B(\phi(\bm{z}^t))$.

It is worth noting that during each online iteration, the Gaussian center positions $\bm{x}_g^t$ are updated synchronously with the deformed mesh $\mathcal{M}(t)$ to ensure strict spatial alignment between the rendering primitives and the underlying geometry (see middle part of Fig.~\ref{fig:pipeline} for illustration). Since the motion of the tetrahedral mesh is governed by a defined physical model, the appearance variations obtained through GS rendering inherently adhere to these physical dynamics. This also showcases the significance of physical simulation, which mitigates artifacts caused by sparse views. Furthermore, the interpolation between positions is differentiable, supporting the chain rule for gradient propagation as discussed in the following section.

\subsection{Window-based Online R2S Adaptation objectives}

For online real-to-sim adaptation, a key challenge lies in balancing the accumulation of historical data and the need for immediate responsiveness. Drawing inspiration from offline methods — where the \textit{dataset length} encompasses the entire interaction sequence — we introduce the concept of a \textit{window} to formulate the online objective. In this context, a \textit{window} represents a rolling buffer of recent observations.
Unlike the offline setting, where the window effectively spans the full history, the online window size is deliberately constrained. This constraint serves a dual purpose: it remains large enough to capture material characteristics and temporal dynamics, yet small enough to ensure computational efficiency for real-time performance and accommodate material variations. Consequently, the windowed differentiable-simulation objective function with vision-based input data is defined as:
\begin{equation}
\mathcal{L}_{\text{r2s}-vision}^{\text{online}}=\sum_{t \in \{T_e - w + 1,\dots,T_e\}}
\sum_{c=1}^K |\mathcal{R}(\mathcal{G}_t) - I_{real}^c|
\label{eq:r2sonline}
\end{equation}
where $\mathcal{R}(\cdot)$ is the differentiable renderer based on 3DGS~\cite{kerbl20233d}, and $\{I_{real}^c\}_{c = 1,2,...K}$ (K is the number of views) are the captured images. By minimizing the online-updated objective function $\mathcal{L}_{\text{r2s}-vision}^{\text{online}}$, the deviation between simulation results and observation data over the window time horizon can be eliminated with updated material properties in the simulation. It is worth noting that a smaller window enables faster parameter updates but is more susceptible to noise, while a larger window leads to slower updates but yields more stable and accurate results - detailed results and discussion in Sec.~\ref{sec:Dis}. 

\begin{figure}[t]
  \centering
  \includegraphics[width=\linewidth]{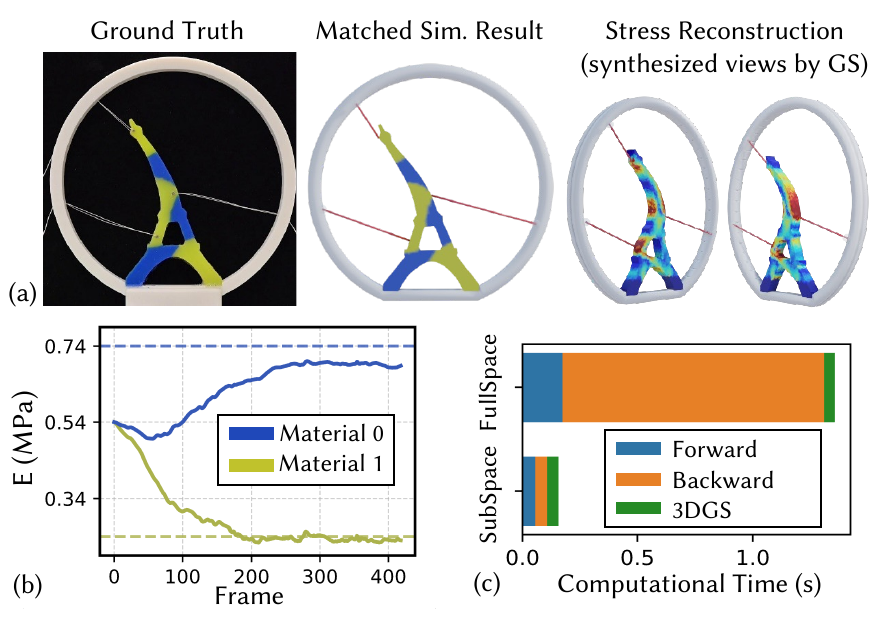}
  \vspace{-20px}
  \caption{Online real-to-sim results for a multi-material cable-driven deformable structure. (a) With vision feedback, the simulation results match the ground truth, enabling downstream applications such as stress reconstruction. (b) Convergence curves for the Young's modulus of two silicone rubber materials. (c) Comparison with computation time showcases that our subspace solution achieves a $7.56\times$ speedup, ensuring online efficiency.}
  \label{fig:tower}
\end{figure}

\begin{figure*}[t]
  \centering
\includegraphics[width=0.9\linewidth]{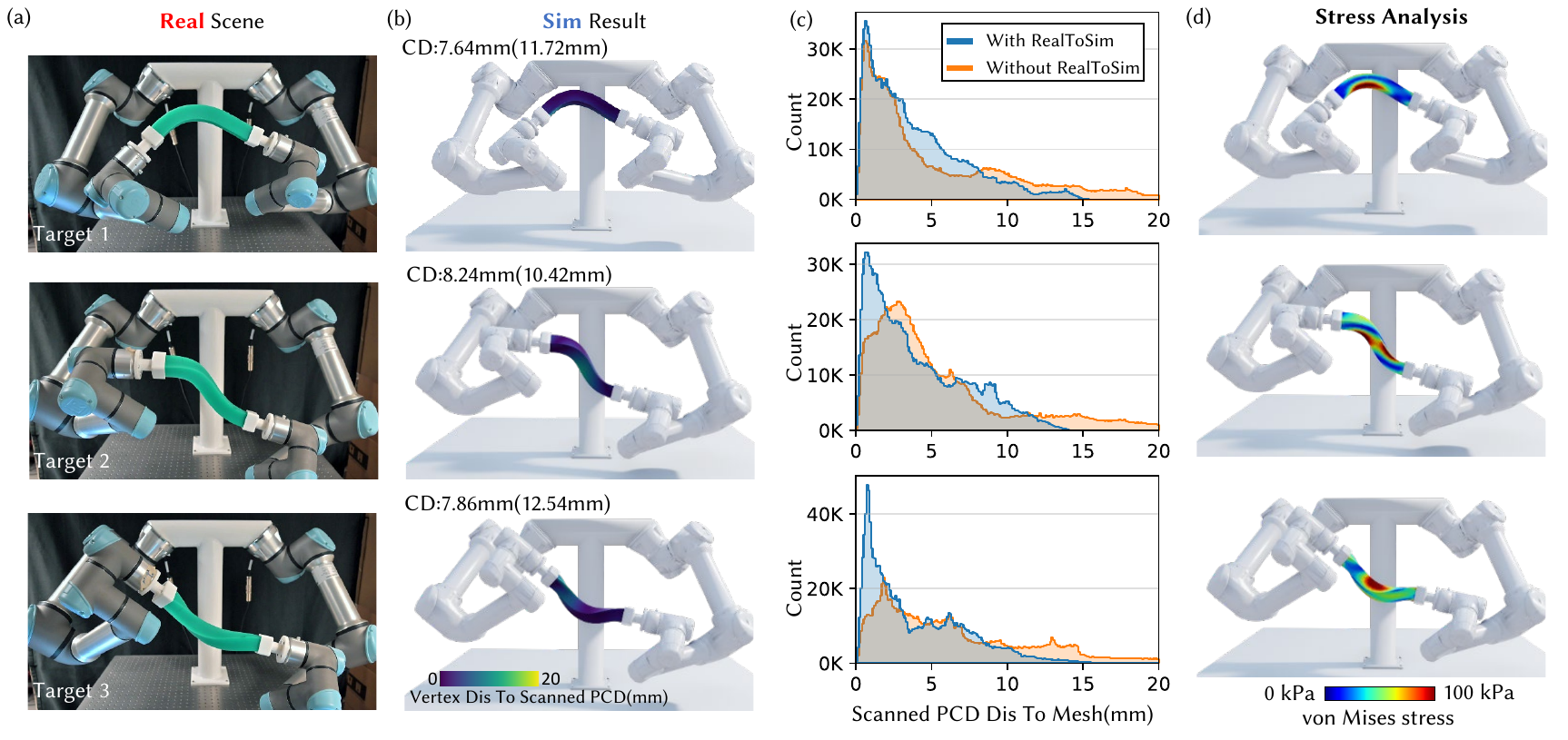}
\vspace{-10px}
  \caption{
   Result for soft bar manipulation by dual robot arms. (a) After performing online real-to-sim adaptation, the control strategy optimized in simulation is directly applied to the real robotic system without additional tuning. (b) Corresponding simulation results after adaptation, colored by the vertex-to-scan distance. (c) Quantitative comparison between the simulated mesh and the ground-truth point cloud captured by a 3D scanner, demonstrating reduced geometric discrepancy after real-to-sim adaptation. (d) The adapted simulation with vision-based feedback further enables downstream physical analysis of von Mises stress reconstruction.
}
  \Description{}
  \label{fig:dualarm-target}
\end{figure*}

\subsection{Gradient-based Optimization}
\label{subsec:optimization}

To achieve online real-to-sim adaptation, we aim to minimize the window-based objective function $\mathcal{L}_{\text{online}}$ (Eq.~\ref{eq:r2sonline}) with respect to the material parameters $\bm{\theta}$ and Gaussian attributes $\mathcal{G}$. 
Mathematically, we aim to minimize the vision-based loss accumulated over the current window $\{T_e - w + 1,\dots,T_e\}$ to update 
the material properties of $\theta$, which follows the formulation of Eq.~\ref{eq:SI_general}:
\begin{equation}
\label{eq:optimization_objective}
\argmin_{\bm{\theta}, \mathcal{G}_t 
} ~\mathcal{L}_{\text{r2s-vision}}^{\text{online}}(\mathcal{G}_t) \quad \text{s.t.} ~\bm{R}(\bm{Z}, \bm{\theta}) = \bm{0},
\end{equation}
where the constraint represents the discrete state equations of the reduced-order simulation (derived from Eq.~\ref{eq:forward_sub}).

While the gradients for Gaussian attributes can be computed directly via the differentiable rasterizer, computing the gradient for material parameters, $d\mathcal{L}/d\bm{\theta}$, is non-trivial due to the complex dependency of the simulation state on $\bm{\theta}$. 
We solve this constrained optimization problem using the adjoint method, which propagates gradients backward through time \rev{}{without explicitly forming the dense Jacobian of the simulation trajectory w.r.t $\bm{\theta}$ and previous states}, while efficiently exploiting system sparsity.

\subsubsection{Vision-to-State Gradient Backpropagation}
First, we compute the gradient of the loss with respect to the reduced simulation state $\bm{z}^t$ at each time step $t$ within the window. Applying the chain rule through the differentiable rendering and the subspace mapping we have
\begin{equation}
\label{eq:vision_grad}
\frac{\partial \mathcal{L}}{\partial \bm{z}^t} = \frac{\partial \mathcal{L}}{\partial I^t} \frac{\partial I^t}{\partial \mathcal{G}^t} \frac{\partial \mathcal{G}^t}{\partial \bm{x}^t} \cdot \bm{J}(\bm{z}^t),
\end{equation}
The first two term aggregates gradients from the pixel level to the GS geometry, while the third term projects these geometric gradients into the reduced simulation subspace.

\subsubsection{Adjoint Method for Reduced-Order Physics}
With the state gradients $\frac{\partial \mathcal{L}}{\partial \bm{z}^t}$ available, we must propagate them to the material parameters $\bm{\theta}$. The relationship between $\bm{z}$ and $\bm{\theta}$ is governed by physics equilibrium. Specifically, we define the physics residual as $\bm{J}(\bm{z}^{t})^\top\bm{r}(\bm{z}^{t}, \bm{y}^{t}, \bm{\theta})$ (see Eq.~\ref{eq:forward_sub}). We seek $\bm{z}^{t}$ such that the residual is orthogonal to the tangent space of the neural subspace mapping:
\begin{equation}
\label{eq:governing_projected}
\bm{h}(\bm{z}^{t}, \bm{z}^{t-1}, \bm{z}^{t-2}, \bm{\theta}) := \bm{J}(\bm{z}^{t})^\top \bm{r}(\bm{z}^{t}, \bm{y}^{t}, \bm{\theta}) = \bm{0}.
\end{equation}
To compute the total derivative $d\mathcal{L}/d\bm{\theta}$, we introduce adjoint variables $\bm{\lambda}_t \in \mathbb{R}^r$. The adjoint variables are computed by solving the following linear system backward in time (from $t=t_e$ down to $t_s$):
\begin{equation}
\label{eq:adjoint_recurrence}
\sum_{k=t}^{t+2}
\left(
\frac{\partial \bm{h}^{k}}{\partial \bm{z}^{t}}
\right)^{\top}
\bm{\lambda}_{k}
=
- \frac{\partial \mathcal{L}}{\partial \bm{z}^{t}} .
\end{equation}
The matrix on the left-hand side is the transpose of the reduced tangent stiffness matrix. Substituting the residual definition, this is efficiently computed as:
\begin{equation}
\begin{aligned}
&\frac{\partial \bm{h}^{t}}{\partial \bm{z}^{t}} = 
\frac{\partial \bm{J}(\bm{z}^{t})^{\top}}{\partial \bm{z}^{t}} \, \bm{r}^{t} 
+ \bm{J}(\bm{z}^{t})^{\top} \frac{\partial \bm{r}^{t}}{\partial \bm{z}^{t}}, \\
&\frac{\partial \bm{h}^{t+1}}{\partial \bm{z}^{t}} = 
\bm{J}(\bm{z}^{t+1})^{\top} \frac{\partial \bm{r}^{t+1}}{\partial \bm{z}^{t}}, ~
\frac{\partial \bm{h}^{t+2}}{\partial \bm{z}^{t}} = 
\bm{J}(\bm{z}^{t+2})^{\top} \frac{\partial \bm{r}^{t+2}}{\partial \bm{z}^{t}} .
\end{aligned}
\end{equation}
where $\partial \bm{J}/\partial \bm{z}^{t}$corresponds to the Hessian of the nonlinear mapping $\phi$.
Finally, once the adjoint variables $\bm{\lambda}$ are computed for the entire window, the gradients with respect to the material parameters are accumulated through the controlled \textit{window}:
\begin{equation}
\frac{d \mathcal{L}}{d \bm{\theta}} = \sum_{t=T_e-w+1}^{T_e} \bm{\lambda}_{t}^\top \left( \frac{\partial \bm{h}^t}{\partial \bm{\theta}} \right) = \sum_{t=T_e-w+1}^{T_e} \bm{\lambda}_{t}^\top \left( -\bm{J}(\bm{z}^{t})^\top \frac{\partial \bm{f}(\bm{z}^{t}, \bm{\theta})}{\partial \bm{\theta}} \right).
\end{equation}
The resulting gradients $\frac{d\mathcal{L}}{d\theta}$ accurately capture how changes in material properties propagate through the reduced subspace together with vision feedback, enabling the optimizer with gradient descent to iteratively refine the physical parameters $\mathbf{\theta}$ at the end of each sliding window, closing the loop between visual observation and physical parameters.

\begin{table*}
\caption{Parameters and results of experiments evaluated in this work. $n$: number of full-space DOFs; $r$: the subspace dimension used; $w$: the window size used; \rev{}{$c$: number of camera views for SI}; $T_r$: average computational time in subspace(s) per frame; $T_f$: average computational time in fullspace(s) per frame.}
\label{table-problems-settings}
\vspace{-10px}
\small
\begin{tabular}{c|c|c|c|c|c|c|c|c|c|c|c}
    \hline
    \multirow{2}{*}{Model} & \multirow{2}{*}{Figure} & \multirow{2}{*}{$n$} & \multirow{2}{*}{$r$} & \multirow{2}{*}{$w$} & \multirow{2}{*}{$c$} & \multicolumn{5}{c|}{Comp. time for \textit{online} SI per frame (s)} & \multirow{2}{*}{Accel. Rate}  \\
    \cline{7-11}
    & & & & & & forward & backward & 3DGS & total $T_r$ & total $T_f$ & \\
    \hline 
    Material Bar & Fig.~\ref{fig:teaser} & 10K & 30 & 3 & 4 & 0.17 & 0.14 & 0.22 & 0.53 & 3.30 & 6.23x \\
    \hline 
    Eiffel tower & Fig.~3, 8 & 12K & 10 & 3 & 2 & 0.06 & 0.07 & 0.05 & 0.18 & 1.36 & 7.56x \\
    \hline 
    Dinosaur & Fig.~4, 7 & 15K & 100 & 5 & 8 & 0.24 & 0.32 & 0.45 & 1.01 & 7.77 & 7.69x \\
    \hline 
    Gingerbread Joy & Fig.~5, 6 & 3.6K & 80 & 15 & 1 & 0.19 & 0.17 & 0.22 & 0.58 & 1.46 & 2.52x \\
    \hline
    High Speed Ball$^\dagger$ & Fig.~\ref{fig-high-speed-ball} & 3.4K & - & 20 & 1 & - & - & - & - & 1.75 & - \\ 
    \hline 
\end{tabular}
\begin{flushleft}\small
 $^\dagger$ For this highly dynamic case, the reduced-order method failed to converge, and only full-space data is reported here. Detailed discussion see Sec.~\ref{sec:Dis}.
\end{flushleft}
\end{table*}

\section{Results and Discussion}

This section details the implementation and results of our \textit{online} real-to-sim adaptation framework using vision feedback. We refer to the supplemental video for detailed results.

\subsection{Implementation and Details of Training}
We implement the differentiable reduced-order simulation module in JAX~\cite{jax2018github} and use the stable neo-Hookean~\cite{smith2018stable} material model to define the elastic energy, while the 3DGS module is implemented in PyTorch~\cite{paszke2017automatic}. For processing physically collected vision data, SAM 2~\cite{ravi2024sam2} is used.
During simulation, the subspace Hessian matrix is obtained via automatic differentiation. 
\rev{}{Additionally, we apply the data-generation procedure by combining different material parameters with randomly generated interaction sequences, allowing the learned subspace to cover diverse soft-body deformations. The material properties (e.g., $E$ and $\nu$), are linearly sampled within the range of physical prior for each frame to create the dataset.}

In this work, all training and experiments are conducted on an NVIDIA RTX 4090 GPU.
The experiments presented in this chapter adopt the AutoEncoder (AE)-based subspace network architecture proposed in~\cite{fulton2019latent}. \rev{}{The comparison with other reduced-order methods (e.g., linear subspace-based representations~\cite{wang2015linear, benchekroun2023skinning}) is presented in Sec.\ref{subsec:comparsion}}. The encoder and decoder of the AE both consist of fully connected hidden layers with a size of $2\times200$, and the learning rate is set to $1\times10^{-3}$. \rev{}{The corresponding subspace dimensions are reported in the table~\ref{table-problems-settings} (ablation study on the selection is discussed in Sec.~\ref{subsec:ablation}.)} To construct the dataset for subspace training, we generate data by combining different material configurations with randomly synthesized interaction sequences. This strategy aims to ensure that the learned subspace adequately covers diverse deformation patterns arising during online state inference.

\subsection{Computation and Physical Results}
\label{subsec:result}

We evaluated the framework across diverse physical scenarios to demonstrate robustness;  Tests involving deformable object interactions highlight the method's efficiency. While the backward pass typically dominates computation time due to high-dimensional Jacobian and Hessian evaluations, our reduced order model significantly accelerates this step, achieving a $2.52 - 7.69\times$ speedup (\rev{}{detailed breakdown of the computational cost can be found in Table~\ref{table-problems-settings}.)}

\begin{figure}[t]
  \centering
  \includegraphics[width=0.8\linewidth]{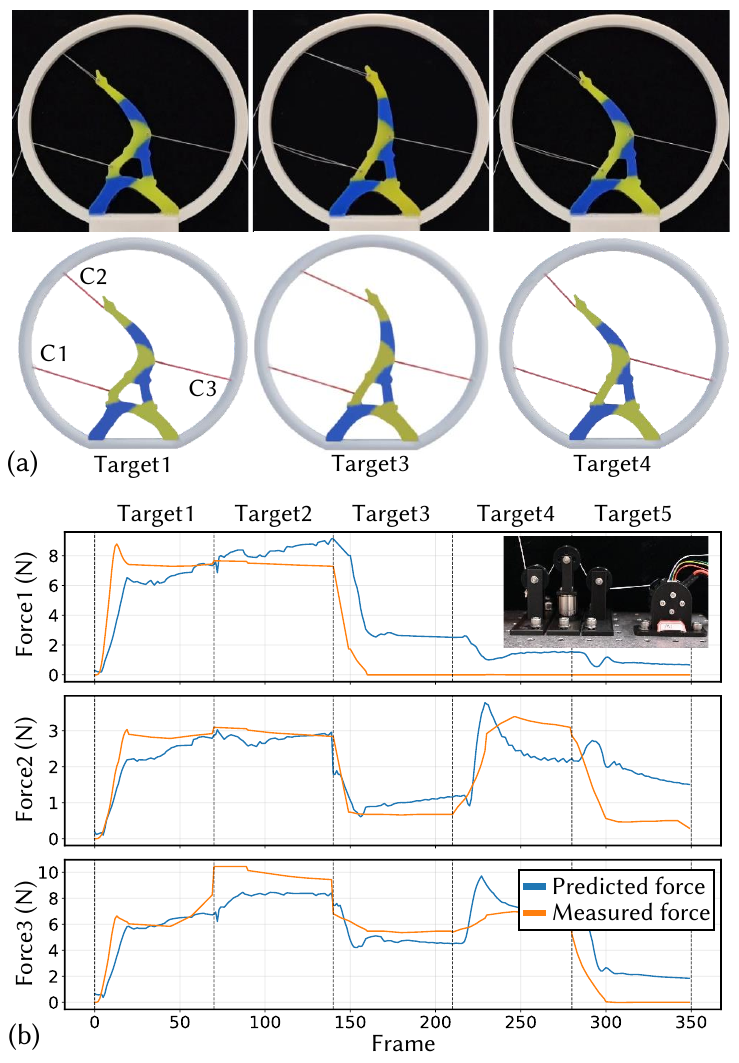}
  \vspace{-10px}
  \caption{(a) Additional real-to-sim adaptation result for cable-driven interaction with the multi-material Eiffel Tower model. (b) With the help of simulation, we can predict the cable force based on stress field reconstruction, which matches the ground-truth data captured by the force sensor, as shown in the zoom-in view. }
  \Description{}
  \label{fig:tower_add}
\end{figure}

The first case we tested is deformable object manipulation by a dual robot arm. As shown in Fig.~\ref{fig:teaser}, positional constraints are applied at both ends of a square-shaped bar made with silicon rubber (material: Smooth-on Dragon Skin 10A). This setup induces pronounced twisting and stretching deformations under the motion of two UR5e robot arms. With the help of the vision feedback from 4 cameras viewed in a stack of window time, the real-to-sim gap is reduced by dynamically updating the material properties. Especially with the reduced-order method, computation time is reduced by 6.23 times (i.e., from 3.30 s to 0.53 s per frame) \rev{}{and the reconstruction error has also decreased to an average of 4.6 mm ($2.3\%$ of the model size).}

With the updated material properties, we further perform a shape-control test using the dual-arm robot setup. The control command is computed using model predictive control (MPC) in simulation with the method presented in~\cite{zhang2025manipulating}, where a dynamics model rolls out candidate dual-arm 6-DoF end-effector action sequences, shape error with respect to the goal state is minimized to optimize the action sequence iteratively, and the first optimized action is executed after inverse-kinematics conversion. As shown in Fig.~\ref{fig:dualarm-target}(a), directly applying the configuration computed in simulation produces physical results that closely match with target shapes, with Chamfer Distance reported in Fig.~\ref{fig:dualarm-target}(b). 

As downstream applications of real-to-sim adaptation, our pipeline recovers internal physical states for stress reconstruction (Figs.~\ref{fig:teaser}(b), \ref{fig:tower}(a), \ref{fig-gingerbread-show}(b)). We calculate the Cauchy stress $\bm{\sigma}$ from the identified parameters and deformation gradient, then derive the Von Mises stress $\sigma_{\mathrm{vM}}$ via the deviatoric stress $\bm{s}$:
\begin{center}
    $\sigma_{\mathrm{vM}} = \sqrt{1.5 \, \bm{s} : \bm{s}}, \quad \text{where } \bm{s} = \bm{\sigma} - \mathrm{tr}(\bm{\sigma}) \bm{I}/3. $
\end{center}
\rev{}{These stresses are transferred to Gaussian attributes to enable real-time, multi-view visualization for digital twins. For example, the stress distribution within the soft bar can be accurately reconstructed at each frame (see Fig.~\ref{fig:dualarm-target}(d)), providing additional physical insight beyond surface geometry. We refer readers to the supplemental video for a better illustration.}

\begin{figure}[t]
  \centering
  \includegraphics[width=0.9\linewidth]{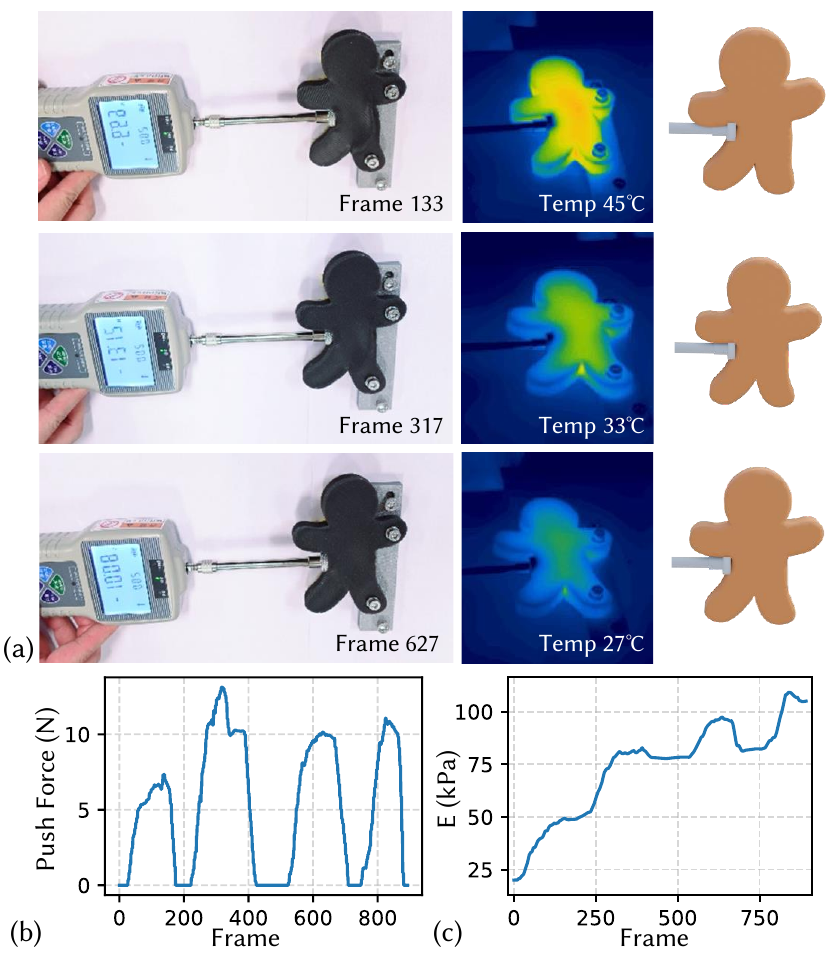}
  \vspace{-10px}
  \caption{
Online real-to-sim adaptation for stiffness tracking in a temperature-varying structure.
(a) Representative frames showing simulated deformation during cooling. (b) Push force over time. (c) Estimated Young’s modulus over time, demonstrating tracking of temperature-dependent stiffness variation.
  }
  \label{fig:gingerbread-real}
\end{figure}

\begin{figure}[t]
  \centering
  \includegraphics[width=0.85\linewidth]{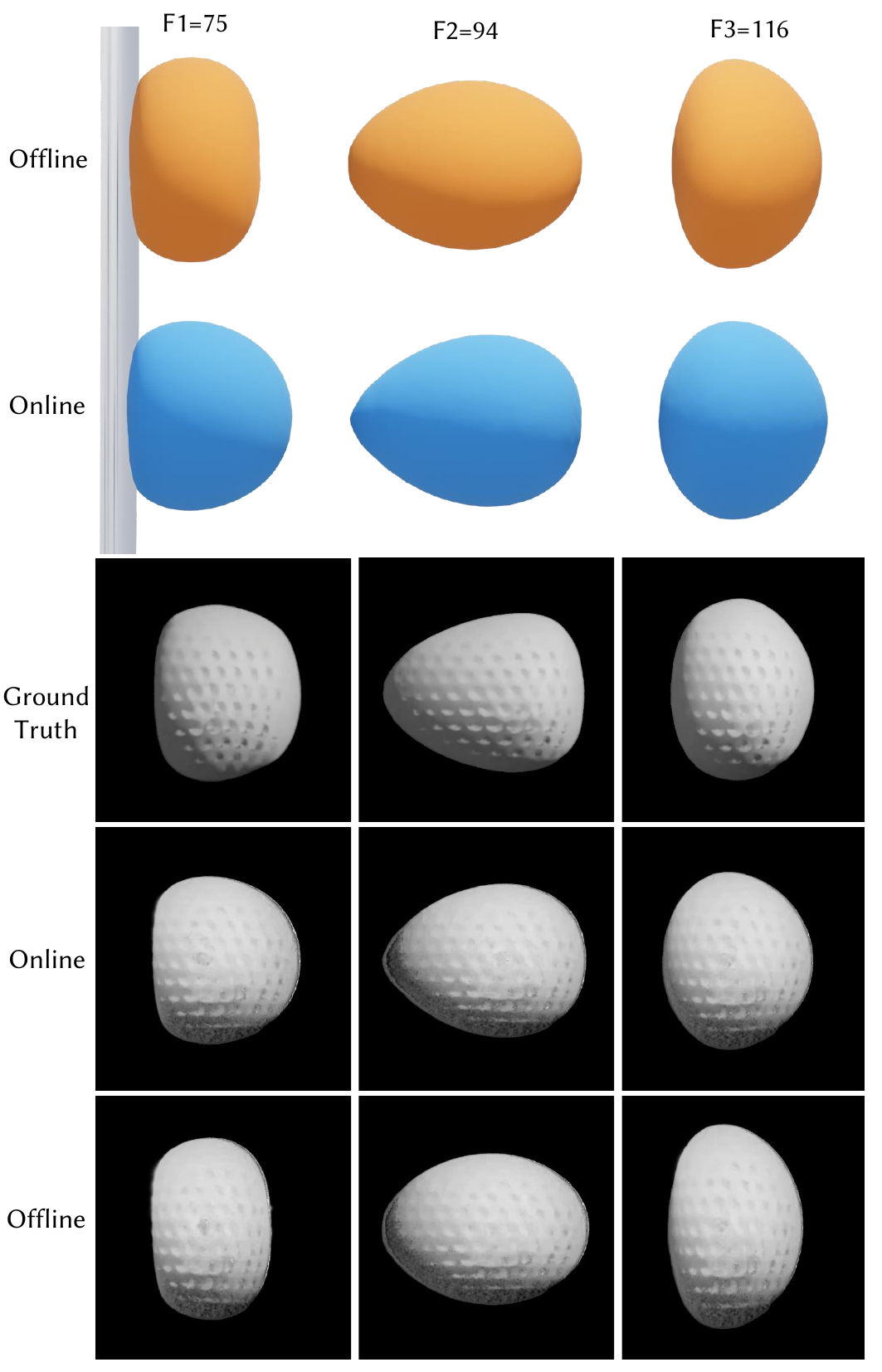}
  \vspace{-10px}
  \caption{Result of a high-speed ball bouncing back after hitting a wall with~$\upsilon_0=\mathrm{27m/s}$. The vision data are duplicated from IPC~\cite{li2020incremental}, and the texture of the result is obtained by GS-based training.
  }
  \Description{}
  \label{fig-high-speed-ball}
\end{figure}

\begin{figure}[t]
  \centering
\includegraphics[width=0.9\linewidth]{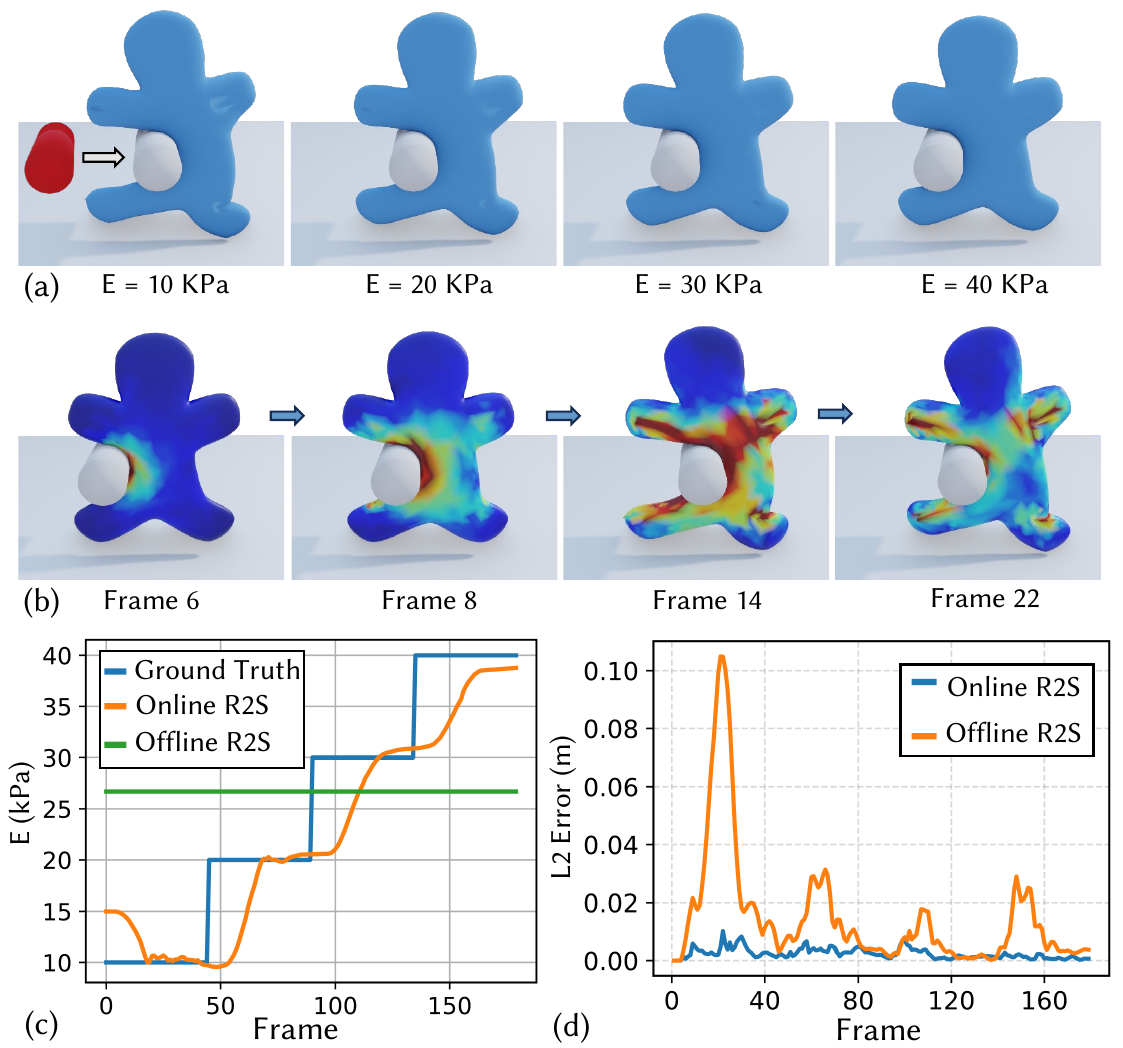}
  \caption{(a) Comparison of deformation dynamics across different stiffness levels (Young's modulus changing dynamically from 10 kPa to 40 kPa in steps when the bar is hit by the Gingerbread-man. (b) Reconstructed stress fields over time (E = 10 kPa for all frames). (c) Prediction of time-varying Young’s modulus changes and (d) comparison of online and offline methods, showing that the proposed online framework tracks evolving stiffness.  }
  \label{fig-gingerbread-show}
\end{figure}

\begin{figure}[t]
  \centering
\includegraphics[width=1\linewidth]{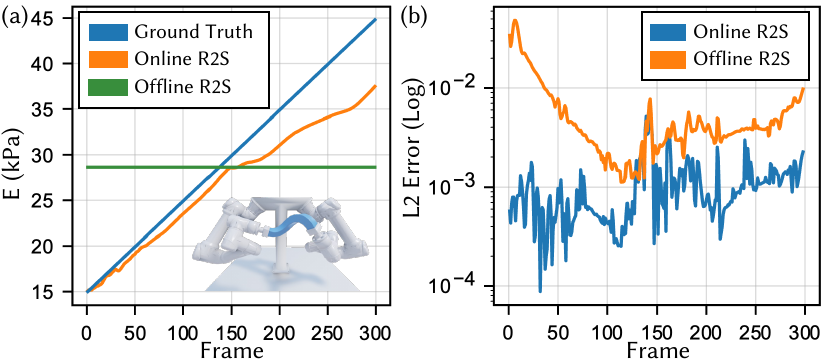}

  \caption{\rev{}{Result and analysis of material aging in the bar manipulation task.
(a) Our online real-to-sim method accurately tracks the time-varying Young’s modulus of the bar, whereas the offline baseline estimates an almost constant stiffness. (b) The per-frame L2
  error further shows that online real-to-sim consistently outperforms the offline method. 
  }
  }
  \label{fig-bar-appendix}
\end{figure}

To verify the method's ability to handle spatially varying stiffness, we utilized a cable-driven interaction setup with the Eiffel Tower model. 
\rev{}{As illustrated in Fig.~\ref{fig:tower} and Fig.~\ref{fig:tower_add}}, the structure is composed of distinct material zones using two silicone materials \rev{}{(Dragon Skin 30A and 10A, with $E = 0.74$ MPa and $0.24$ MPa, respectively). With the proposed pipeline, multiple independent elasticity parameters are accurately identified. Similarly, the stress distribution during cable actuation can be reconstructed, as shown in Fig.~\ref{fig:tower}(a), which further enables external cable-force prediction by computing internal nodal forces from the reconstructed stress field.} For each cable, we identify the attached vertices and approximate the total tension by summing the magnitudes of the internal forces at those vertices. \rev{}{As shown in Fig.~\ref{fig:tower_add}(b), the computed cable force closely matches the force measured by the sensor in the physical setup.}  

We also demonstrate the ability of the proposed online real-to-sim adaptation in stiffness tracking for a temperature-varying structure. As shown in Fig.~\ref{fig:gingerbread-real}, the gingerbread-man-shaped specimen was fabricated using a 3D-printed polycaprolactone (PCL) truss structure embedded within Ecoflex 00-30 silicone through a molding process. With temperature decreases from 45$^\circ$C to 27$^\circ$, it progressively increases the stiffness of the PCL structure~\cite{baptista2020effect}, resulting in changes in deformation and force responses under similar interactions. Our method continuously adapts online real-to-sim adaptation, and well captures the increasing push force and estimated increasing Young's modulus. 

\begin{figure}[t]
  \centering
  \includegraphics[width=0.9 \linewidth]{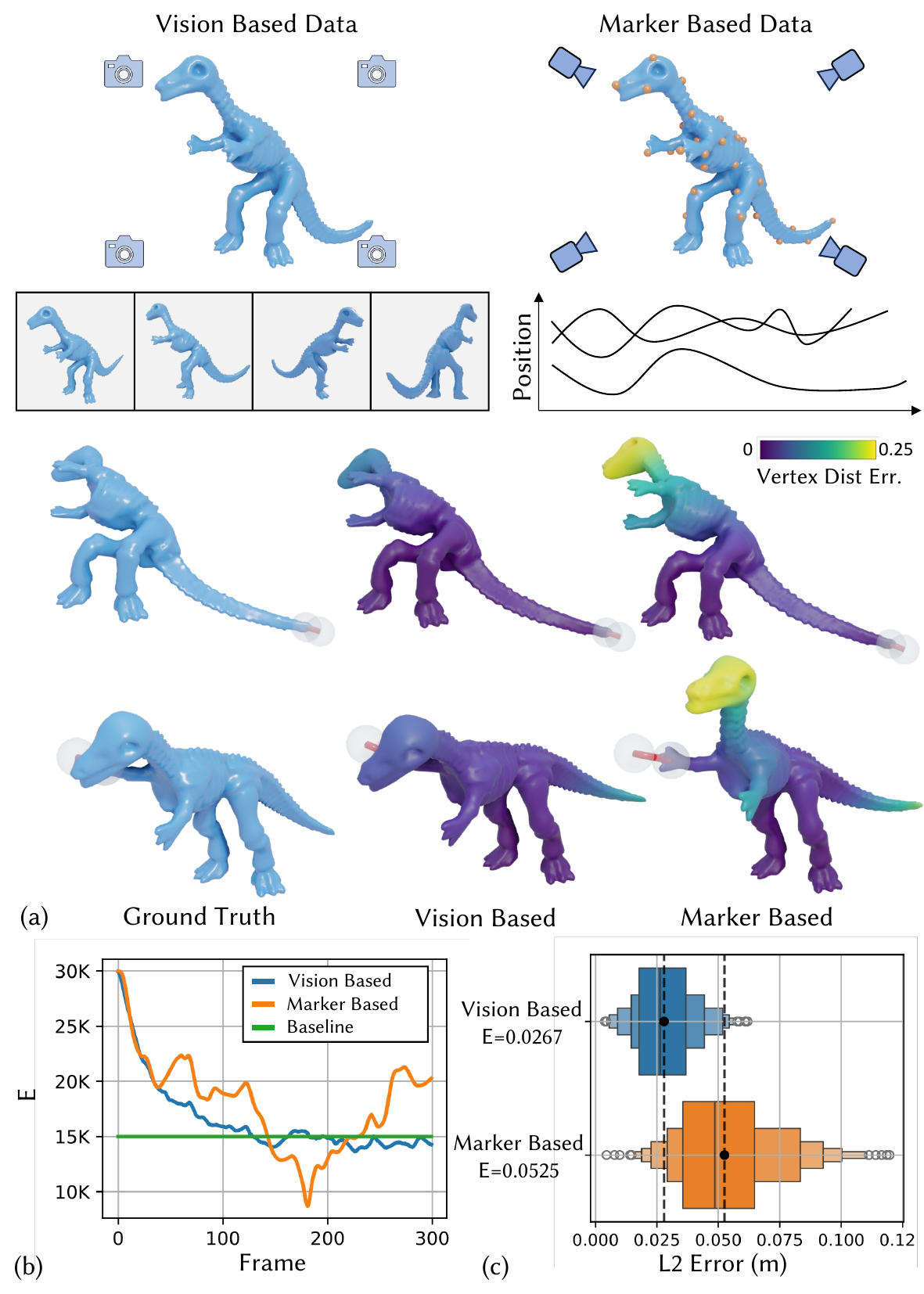}
  \vspace{-10px}
  \caption{Comparison of online real-to-sim adaptation on the dinosaur example using vision-based and marker-based physical input. 40 markers uniformly distributed are used for comparison. (a) Key frame showcase vision-based method outperforms in terms of reconstructed geometry accuracy. (b) Material parameter update curves during \textit{online} adaptation. (c) Distribution of L2 errors over all frames for the two methods.
  }
  \label{fig-dinosaur-show}
\end{figure}

Validation cases involving collisions and high-frequency dynamics are also tested. This includes the golf-ball bouncing case, our online method better matches the ground-truth transient deformation, while the offline baseline shows noticeable shape errors during compression and rebound - \rev{}{as a comparison, the PSNR greatly increased from 18.0 to 22.6 with our online method for the first key frame shown in Fig.~\ref{fig-high-speed-ball}}. For ease of comparison with existing work and to conduct ablation studies, we also include a virtual test case using the Dinosaur model with random external interactions (see Fig.~\ref{fig-dinosaur-show}), and details of this case are discussed in the next section.

\subsection{\rev{}{Comparison, Ablation Study, and Discussion}}
\label{sec:Dis}

\subsubsection{Comparison with existing offline method.}  \label{subsec:comparsion}

\rev{}{We compare our online real-to-sim approach with existing vision-based offline methods~\cite{hahn2019real2sim, chen2025vid2sim}.} Unlike our method, the offline baseline estimates a single fixed model from the full sequence. This strategy is less effective for long-horizon interactions, especially when the system exhibits time-varying materials or accumulates modeling errors over time. A detailed virtual comparison is provided in Fig.~\ref{fig-gingerbread-show}(c), where the offline method can failed to capture the material variation and cannot solve for the deformation. \rev{}{We further observe that the offline method remains less effective even when the material properties are fixed. In the two-robot-arm bar manipulation example, our method achieves a higher PSNR of 25.91 compared with 24.42 for the offline baseline. This suggests that optimizing a limited set of material parameters offline is insufficient to fully capture the physical behavior in long-horizon manipulation tasks, where accumulated errors and unmodeled dynamics can degrade reconstruction accuracy.}

\rev{}{A similar performance difference is observed in the material aging experiment. As shown in Fig.~\ref{fig-bar-appendix}(a), our online method closely tracks the increasing stiffness of the bar over time, whereas the offline baseline converges to an almost constant stiffness and fails to capture the temporal variation. Consequently, the offline method overestimates the stiffness in the early frames and underestimates it in the later frames as the ground-truth stiffness increases. The per-frame \(L_2\) error in Fig.~\ref{fig-bar-appendix}(b) further shows that our method consistently achieves lower simulation error, reducing the loss by approximately \(3\text{--}5\times\) compared with the offline baseline.}

\begin{figure}[t]
  \centering
\includegraphics[width=0.65\linewidth]{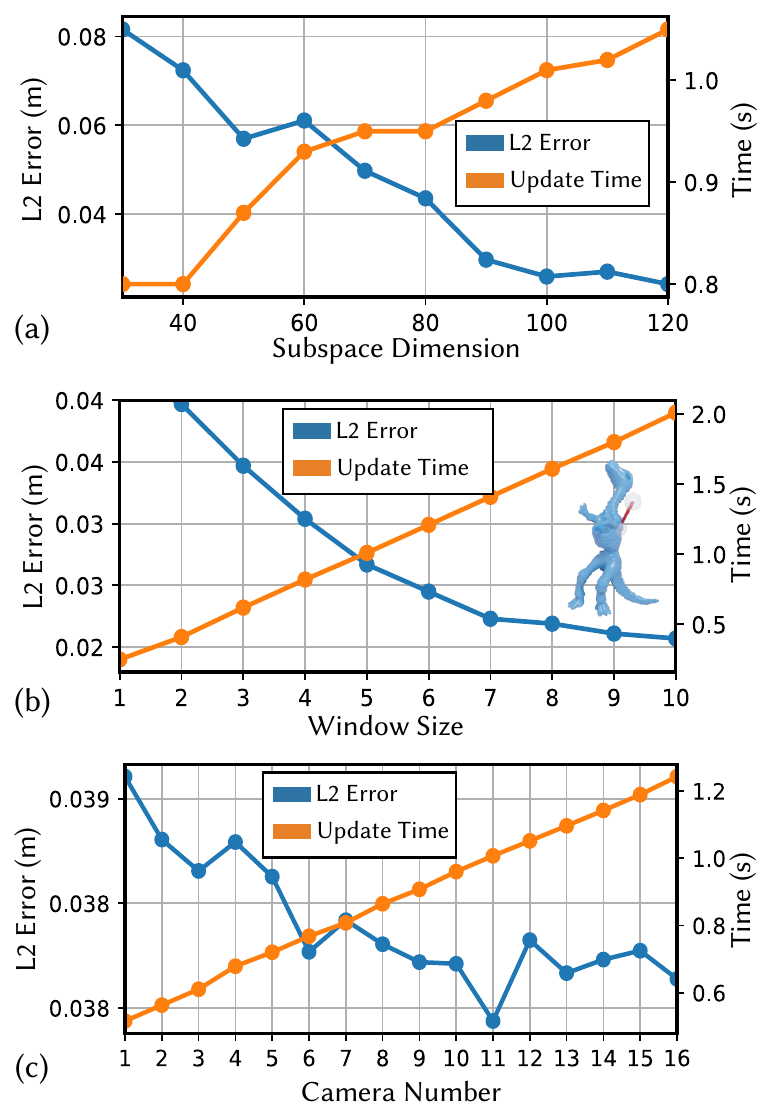}
\vspace{-10px}
  \caption{\rev{}{Ablation studies on the Dinosaur model with random interactions.
  (a) Increasing the subspace dimension reduces the L2 error but increases the update time.
(b) A larger temporal window size generally improves accuracy at the cost of longer update time (for this test, a fixed $r = 100$ is used). (c) Ablation on the number of camera views $c$ for both reconstruction error and computational cost.}}
  \label{fig-ablation}
\end{figure}

\subsubsection{Ablation Study on Parameter Selection}
\label{subsec:ablation}
We conduct an ablation study on hyperparameter selection with the dinosaur model under random contact conditions (Fig.~\ref{fig-ablation}), focusing on the subspace dimension $r$, temporal window size $w$, \rev{}{and number of camera view $c$}. It can be seen that increasing $r$ improves deformation expressiveness and rapidly reduces the L2 reconstruction error from $r=30$ to $r=90$, after which the error plateaus, while update time grows approximately linearly. We choose $r=100$ as a trade-off point, where the error stabilizes around 0.025~m, i.e., about $5\%$ of the model size. With $r=100$ fixed, increasing $w$ incorporates more temporal constraints and further reduces the error, but also linearly increases update time; the improvement becomes marginal after $w=5$. \rev{}{We also present the ablation of camera number in Fig.~\ref{fig-ablation}(c), where can be seen that increasing the number of cameras initially reduces the reconstruction error, indicating that additional views provide useful constraints for the online update. However, once the multi-view images cover most of the geometry, the accuracy improvement becomes marginal, while the update time continues to increase approximately linearly. } Therefore, $r$, $w$, and $c$ are tuned for each model to balance reconstruction accuracy and computational efficiency - we refer Table~\ref{table-problems-settings} for statistics.

\subsubsection{Discussion on Constitutive Model Choice and Vision-based Feedback.}
\begin{figure}[t]
  \centering
\includegraphics[width=0.9\linewidth]{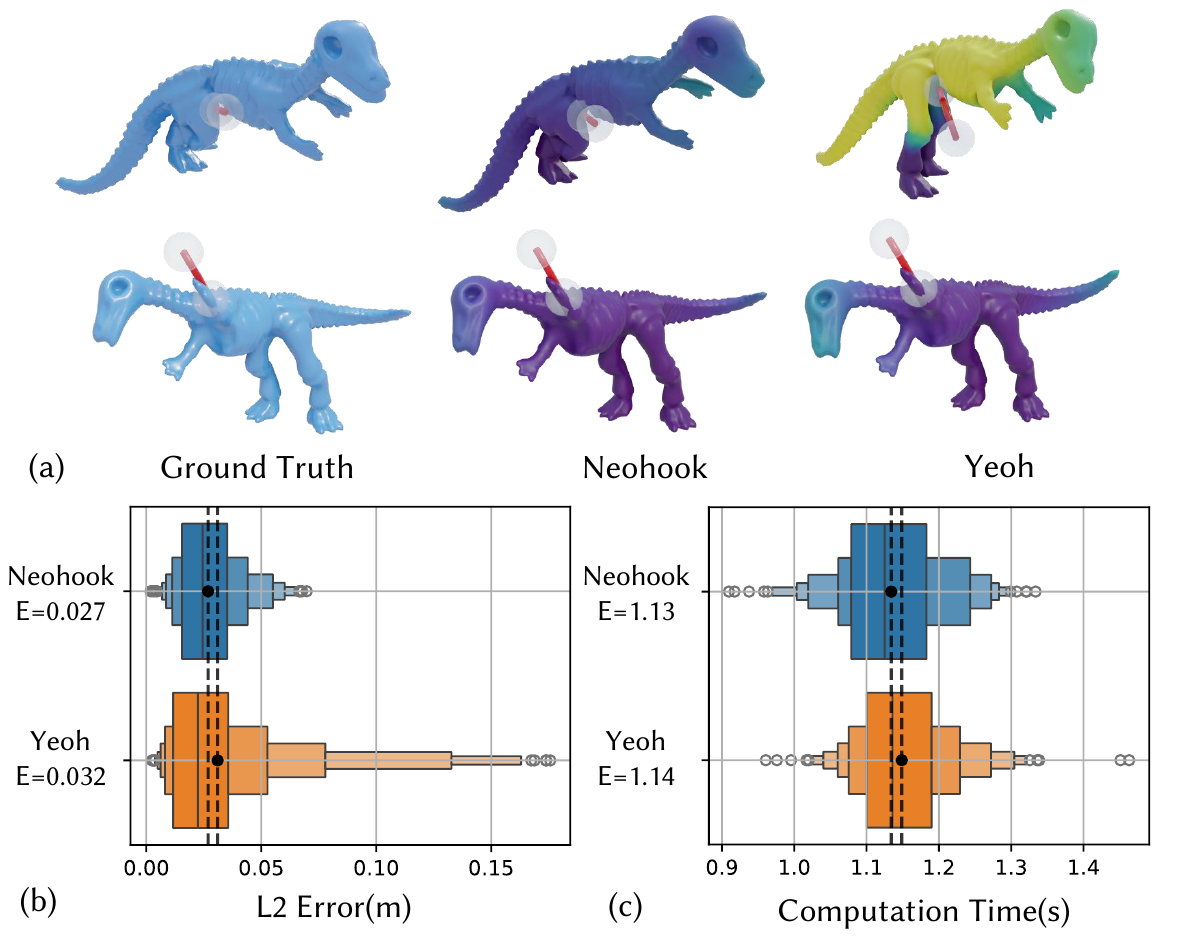}
  \vspace{-10px}
  \caption{
Comparison of different material models tested with our pipeline. (a) Qualitative comparison between the ground-truth geometry and the adapted simulation results using the two material models. (b, c) With the proposed method, the Yeoh model shows similar computation time but less robust real-to-sim adaptation under large deformation.
  }
  \Description{}
  \label{fig:dinosaur-multi-material}
\end{figure}

We further examine how the choice of constitutive model affects online real-to-sim adaptation. As shown in Fig.~\ref{fig:dinosaur-multi-material}, although the Yeoh model provides a higher-order formulation with four material parameters, it is less robust under large deformation and produces larger reconstruction errors than the Neo-Hookean model. This suggests that a more expressive material model does not necessarily improve online adaptation, as the higher-dimensional parameter space can make optimization less stable from limited observations. Both models require similar online update time within our reduced-order differentiable pipeline, indicating that the framework remains efficient even with the higher-dimensional Yeoh parameterization. \rev{}{On the other hand, we also compare with conventional learning based methods~\cite{wang2015linear,benchekroun2023skinning} to demonstrate the effectiveness of the neural subspace selection. In the same Dinosaur setting, the average reconstruction error is decreases from 24.34 to 20.15 with our AutoEncoder-based subspace network~\cite{fulton2019latent} reduces it to 20.15, demonstrates that it can better capture complex nonlinear deformation modes than classic Linear Blend Skinning (LBS) representations~\cite{chen2025vid2sim} for the presented real-to-sim tasks.} 

Additionally, the proposed online real-to-sim adaptation framework supports both marker-based and vision-based observation inputs. We compare the two on the bar and dinosaur manipulation cases and find that vision-based feedback is more effective. As shown in Fig.~\ref{fig-dinosaur-show}(c), the average vertex $L_2$ error decreases from 0.052~m to 0.026~m, while the material update curves in Fig.~\ref{fig-dinosaur-show}(b) show faster convergence due to the denser spatial information provided by images. Similarly, Fig.~\ref{fig:teaser}(d) shows that the vision-based solution reduces vertex error by about $50\%$, substantially reducing the real-to-sim error.


\subsubsection{Limitation and Future Work}
While our framework is robust across diverse scenarios, several limitations remain. Although we show effective online SI for multi-material setups with clearly defined regions, extending to per-element material distributions would greatly increase the parameter space and complicate optimization. Addressing this may require stronger regularization or neural representations for continuous material variation. \rev{}{Additionally, reduced-order models may face convergence issues in highly dynamic systems and have limited extrapolation capability for unseen deformation modes or out-of-range material parameters~\cite{sharp2023data,lyu2024accelerate}.} Therefore for such cases, e.g., the high-speed ball in Fig.~\ref{fig-high-speed-ball}, we revert to full-space simulation. Even without model reduction, our online framework remains more accurate than the offline method, producing simulations that better match visual observations. \rev{}{Future work will explore adaptive refinement to retain reduced-order efficiency in highly dynamic cases, and inviting unsupervised learning to improve the extrapolation ability.}

\section{Conclusion}
In this work, we present an online real-to-sim adaptation framework that integrates differentiable reduced-order simulation with visual feedback. By combining a neural subspace solver with window-based optimization, our method alleviates the computational bottlenecks of \textit{offline} full-space methods while capturing time-varying dynamics to reduce sim-to-real gaps. Extensive virtual and physical experiments on elastic object manipulation, cable-driven multi-material manipulation, and temperature-dependent stiffness tracking show the effectiveness of our pipeline. Our method consistently outperforms offline baselines and marker-based approaches, supporting downstream applications such as force prediction, 3D stress reconstruction, and robot-based manipulation. We believe this capability opens new opportunities for building digital twins that mimic the physical world with dynamic interactions.

\begin{acks}
The authors would like to thank Mr. Chenyu Xu for his kind help in conducting time-varying material experiments, and Mr. Aoran Lyu for insightful discussions during the early stages of this work.

This work is supported by the Guangdong Basic and Applied Basic Research Foundation (2025A1515010124), and in part by the National Key Research and Development Program of China (2024YFF1206600), the National Natural Science Foundation of China (62325204), the HKSAR Research Grants Council Early Career Scheme (RGC-ECS) (CUHK/24204924), and in part by the Multi-scale Medical Robotics Center, AIR@InnoHK.

\end{acks}

\bibliographystyle{ACM-Reference-Format}
\bibliography{bibfile}

\end{document}